\documentclass[pra,twocolumn,amsmath,amssymb,groupedaddress,floatfix,showpacs,showkeys,subeqn]{revtex4-1} 
\usepackage{hyperref}
\usepackage{graphics} 
\usepackage{graphicx} 
\usepackage{subfigure}
\usepackage{url}

\newcommand{\sech}{\, \mathrm{sech} \,}
\newcommand{\field}[1]{\mathbb{#1}} 

\begin{document}
\author{L. A. Toikka}
\author{J. Hietarinta}
\author{K-A. Suominen}
\affiliation{Department of Physics and Astronomy, University of Turku, 20014 Turku, Finland}
\title{Exact Soliton-like Solutions of the Radial Gross-Pitaevskii Equation}
\date{\today}
\begin{abstract}
  We construct exact ring soliton-like solutions of the cylindrically
  symmetric (i.e., radial) Gross-Pitaevskii equation with a potential,
  using the similarity transformation method. Depending on the choice
  of the allowed free functions, the solutions can take the form of
  stationary dark or bright rings whose time dependence is in the
  phase dynamics only, or oscillating and bouncing solutions, related
  to the second Painlev\'e transcendent. In each case the potential can be
  chosen to be time-independent.
\end{abstract}
\keywords{ring soliton, nonlinear Schr\"{o}dinger equation, Gross-Pitaevskii equation, cylindrical symmetry, Painlev\'e transcendent}

\maketitle

\section{Introduction}
Many nonlinear equations of motion allow interesting solutions such as bright and dark solitons, which propagate without dispersion~\cite{Drazin1989,Dauxois2006}. More generally, we can also consider localised soliton-like stationary solutions, which have the appearance of solitons except for some aspects of dynamics. They are usually accompanied by a spatially inhomogeneous external potential-like term in the equation of motion and often called also solitons, or solitary waves, or for dark soliton cases, kink-like solutions~\cite{Kinks2006}. 

The nonlinear Schr\"odinger equation (NLS), 
\begin{equation}
\label{eqn:nls}
i\psi_{t} = -\psi_{xx} + V(x,t) \psi + \sigma
|\psi|^2 \psi 
\end{equation}
also known as the Gross-Pitaevskii equation (GPE), is of special interest since it describes the relevant physics in nonlinear optical systems~\cite{Kivshar1998} as well as in degenerate quantum gases~\cite{P&S,Emergent}. For $V=0$ and $\sigma=-1$ it is called focusing and allows bright solitons, while for $\sigma=+1$ it is called de-focusing and allows dark solitons. These solutions tend to work well also when a potential term $V$ is added to the nonlinear equation. Equation~\eqref{eqn:nls} is one-dimensional, and while the trivial extension to two dimensions by replacing the $x$-derivatives by a Laplacian is possible, the corresponding solutions are plane waves without solitonic properties~\cite{Kinks2006}. Typically they are not stable: the stationary bright soliton solution survives only under certain conditions and if supported by an external potential~\cite{Dodd1996}, and the stationary dark soliton decays into vortices by the snaking instability unless the system is in practice a quasi-one-dimensional
one~\cite{PhysRevA.60.R2665}. This applies especially to any static solution of Eq.~\eqref{eqn:nls}, which gives the motivation to seek such multidimensional solutions that can be expected to have at least extended lifetimes. The static soliton-like solutions are also interesting more generally, connecting to kinks and domain walls~\cite{Kinks2006}, although e.g.~in superfluid He the multicomponent structure of the order parameter leads to a rich set of complex structures~\cite{Volovik2003}.

Instead of making such a plane-wave extension into two dimensions, one
can look at cylindrically symmetric systems, which brings forward the
concept of ring solitons. Since their introduction in a nonlinear
optics setting~\cite{PhysRevE.50.R40}, the interest has been in dark
ring solitons, which have been treated essentially numerically and in
the domain of a large
radius~\cite{PhysRevLett.90.120403,Greekreview}. Although one can
reduce the dynamics into a one-dimensional radial equation, exact
analytic solutions (either bright or dark) have not been obtained so
far. It has been only shown that in the limit of infinitesimal
amplitude, such solitons are analytically described by the radial
KdV-equation~\cite{PhysRevE.50.R40}. Apart from being simply
convenient, exact solutions would allow one to analyse the stability
and decay of ring solitons, including the effect of external
potentials, in the same fashion as for trivially extended plane-wave
solutions in two dimensions.

We have found that using a similarity transformation method, one can
actually construct exact analytical ring soliton-like solutions of the
radial Gross-Pitaevskii equation. They are limited to certain
forms of external potential, but even then, they can provide a
starting point for further studies of the ring solitons in
cylindrically symmetric potentials. In this paper we first describe
the similarity transformation in Sec.~\ref{sec:simtf}, and then
present the solutions in Sec.~\ref{sec:RSS}. Then, in
Sec.~\ref{sec:p22} we discuss the similarity solutions associated with
the second Painlev\'e equation. We summarise our work and
discuss its implications in Sec.~\ref{sec:summary}.

\section{\label{sec:simtf}The Similarity Transformation}
The radial GPE in dimensionless form is given by
\begin{equation}
\label{eqn:2}
i\psi_{T} = -\psi_{RR} - \frac{1}{R} \psi_{R} + V(R, T) \psi + \sigma |\psi|^2 \psi
\end{equation}
where $\psi$ is the complex amplitude of the
electric field in a nonlinear optics setting or the macroscopical
wavefunction of a Bose-Einstein condensate in an ultracold atomic gas
setting. In our approach we focus on the latter framework, in which
case $V(R, T)$ is the external potential. We will mainly consider the
defocusing case $\sigma = 1$. We note that Eqs.~\eqref{eqn:nls} and~\eqref{eqn:2} are not connected by a general transformation combining
point-, gauge- or scale-transformations, which were discussed, e.g.,  in
Refs.~\cite{0305-4470-26-23-043}, and~\cite{PhysRevLett.98.064102}.

The variable $R$ in Eq.~\eqref{eqn:2} is strictly positive, so to simplify the treatment it is useful to apply the transformation $x = \ln{(R)}$ and $t = T$, which results in
\begin{equation}
\label{eqn:4}
i\psi_t = -\frac{1}{e^{2x}} \psi_{xx} + V(x, t) \psi + |\psi|^2 \psi.
\end{equation}
From now on we will consider the generalised equation
\begin{equation}
\label{eqn:5}
i\psi_t = -\alpha_1(x,t) \psi_{xx} + \alpha_2(x,t) \psi + \alpha_3(x,t)|\psi|^2 \psi,
\end{equation}
where $\alpha_{1-3}$ are some functions of $x$ and $t$, and consider our case, Eq.~\eqref{eqn:2}, as the following special case:
\begin{equation}
\label{eqn:6}
\alpha_1(x,t) = \frac{1}{e^{2x}}, \qquad \alpha_2(x,t) = V(x, t), \qquad \alpha_3(x,t) = 1.
\end{equation}

We now use the similarity ansatz~\cite{Bluman&Cole,PhysRevLett.100.164102,Zhenya20104838}
\begin{equation}
\label{eqn:5.1}
\psi(x,t) = \rho(x,t)e^{i \varphi(x,t)} \phi(\eta(x,t)),
\end{equation}
where $\rho \in \field{R}^+$ and $\{\varphi, \eta\} \in \field{R}$ are
some functions to be determined, and $\phi(\eta)$ is assumed to satisfy
\begin{equation}
\label{eqn:5.2}
-\phi_{\eta \eta} + g(\eta)\phi_{\eta} + h(\eta)\phi + G\phi^3 = 0,
\end{equation}
where $G$ is a constant and $g$ and $h$ are arbitrary functions of $\eta$. By a simple transformation we can set $g(\eta) = 0$ without loss of generality. We have, for example, the following special cases:
\begin{equation}
\tag{Painlev\'e II}
-\phi_{\eta \eta} + \eta \phi + 2\phi^3 = 0
\end{equation}
with $g=0, h=\eta, G=2$~\cite{NISTPII} and
\begin{equation}
\tag{Canonical NLS}
-\phi_{\eta \eta} + \phi^3 = \mu \phi
\end{equation}
with $g=0, h=-\mu, G = 1$.

Substituting Eq.~\eqref{eqn:5.1} in Eq.~\eqref{eqn:5} we obtain the following set of equations:
\begin{subequations}
\begin{align}
\label{eqn:5.4a}
(\rho^2 \eta_x)_x &= 0, \\
\label{eqn:5.4b}
(\rho^2)_t + 2\alpha_1(\rho^2 \varphi_x)_x &= 0, \\
\label{eqn:5.4c}
\eta_t + 2\varphi_x \eta_x \alpha_1 &= 0, \\
\label{eqn:5.4d}
-\alpha_3 \rho^2 + G \eta_x^2 \alpha_1 &= 0, \\
\label{eqn:5.4e}
\alpha_1(\varphi_x^2 -h \eta_x^2 - \frac{\rho_{xx}}{\rho}) + \alpha_2 + \varphi_t &= 0.
\end{align}
\end{subequations}

Assuming Eq.~\eqref{eqn:6}, we obtain from Eqs.~\eqref{eqn:5.4a}-\eqref{eqn:5.4d} the following solutions:
\begin{align}
\label{eqn:5.100}
\rho &= \frac{e^{-\frac{x}{3}}}{ c_1(t)}, \\
\label{eqn:5.101}
\varphi &= \frac{3 \dot{c}_1(t)}{8c_1(t)}e^{2x} + c_2(t),\\
\label{eqn:5.102}
\eta &= \frac{3}{2\sqrt{G}c_1(t)} e^{\frac{2x}{3}} + c_3,
\end{align}
where $c_{1,2}(t)$ are arbitrary functions of time and $c_3$ is a constant. From Eq.~\eqref{eqn:5.4e} we obtain an expression for the potential:
\begin{equation}
\label{eqn:pot1}
\alpha_2  = \frac{1}{9 e^{2x}} + \frac{h(\eta)e^{-\frac{2x}{3}}}{Gc_1^2(t)} - \frac{3}{16} \frac{\dot{c}_1^2(t) + 2c_1(t)\ddot{c}_1(t)}{c_1^2(t)}e^{2x} - \dot{c}_2(t).
\end{equation}
We will next consider the potentials and solutions that can be obtained with various choices of $c_{1,2}(t)$ and $h(\eta)$.

\section{The Ring Soliton-like Solutions}
\label{sec:RSS}

There is freedom in choosing the potential function $\alpha_2$. For example, the $1/(9R^2)$ term can be removed if we choose $h(\eta) = -\frac{1}{4\eta^2}$ and $c_3 = 0$. Then from Eq.~\eqref{eqn:pot1} we obtain a harmonic potential
\begin{equation}
\label{eqn:varpot1}
\alpha_2  = - \frac{3}{16} \frac{\dot{c}_1^2(t) + 2c_1(t)\ddot{c}_1(t)}{c_1^2(t)}e^{2x} - \dot{c}_2(t).
\end{equation}
See Table \ref{tab:1} for a selection of choices for $h(\eta)$.

\begin{table*}
\caption{A range of potentials available by suitable choices of $h(\eta)$. From Eq.~\eqref{eqn:5.2} we then have the equation for $\phi(\eta)$, which we must solve to construct the solutions $\psi(x,t)$.}
\label{tab:1}
\begin{ruledtabular}
\begin{tabular}{c c l r}
$h(\eta)$ & $c_3$ & Potential & Eq. to solve \\
\hline
$h(\eta)$ & $c_3$ & $\frac{1}{9 R^2} + \frac{h(\eta)}{Gc_1^2(T)R^{\frac{2}{3}}} - \frac{3}{16} \frac{\dot{c}_1^2(T) + 2c_1(T)\ddot{c}_1(T)}{c_1^2(T)}R^2 - \dot{c}_2(T)$ & $-\phi_{\eta \eta} + h(\eta)\phi + G\phi^3 = 0$ \\
$\eta$ & 0 & $\frac{1}{9 R^2} - \frac{3}{16} \frac{\dot{c}_1^2(T) + 2c_1(T)\ddot{c}_1(T)}{c_1^2(T)}R^2 + \frac{3}{2G^{\frac{3}{2}} c_1^3(T)} - \dot{c}_2(T)$ & $-\phi_{\eta \eta} + \eta \phi + G\phi^3 = 0$ \\
$\eta^{\frac{5}{2}}$ & 0 & $\frac{1}{9 R^2} - \frac{3}{16} \frac{\dot{c}_1^2(T) + 2c_1(T)\ddot{c}_1(T)}{c_1^2(T)}R^2 + \left(\frac{3}{2\sqrt{G}c_1(T)}\right)^{\frac{5}{2}} \frac{1}{Gc_1^2(T)} R - \dot{c}_2(T)$ & $-\phi_{\eta \eta} + \eta^{\frac{5}{2}}\phi + G\phi^3 = 0$ \\
$- \frac{1}{4\eta^2}$ & 0 & $- \frac{3}{16} \frac{\dot{c}_1^2(T) + 2c_1(T)\ddot{c}_1(T)}{c_1^2(T)}R^2 - \dot{c}_2(T)$ & $-\phi_{\eta \eta} - \frac{1}{4\eta^2}\phi + G\phi^3 = 0$ \\
$- \frac{1}{4\eta^2}+\eta^{\frac{5}{2}} + \eta $ & 0 & $- \frac{3}{16} \frac{\dot{c}_1^2(T) + 2c_1(T)\ddot{c}_1(T)}{c_1^2(T)}R^2 + \left(\frac{3}{2\sqrt{G}c_1(T)}\right)^{\frac{5}{2}} \frac{1}{Gc_1^2(T)} R$& \\ 
& & 
$+ \frac{3}{2G^{\frac{3}{2}} c_1^3(T)} - \dot{c}_2(T)$ & $-\phi_{\eta \eta} +\left( - \frac{1}{4\eta^2}+\eta^{\frac{5}{2}} + \eta\right)\phi + G\phi^3 = 0$ \\
$-\mu$ & $c_3$ & $\frac{1}{9 R^2} + \frac{-\mu}{Gc_1^2(T)R^{\frac{2}{3}}} - \frac{3}{16} \frac{\dot{c}_1^2(T) + 2c_1(T)\ddot{c}_1(T)}{c_1^2(T)}R^2 - \dot{c}_2(T)$ & $-\phi_{\eta \eta} -\mu \phi + G\phi^3 = 0$ \\
\end{tabular}
\end{ruledtabular}
\end{table*}

\subsection{The dark ring soliton-like solution, $h(\eta) = -\mu$}
\label{sec:kink}

This choice gives the canonical NLS (no external potential) so that from Eq.~\eqref{eqn:5.2} we are now requiring $\phi(\eta)$ to satisfy
\begin{equation}
\label{eqn:1.4}
\mu \phi = -\phi_{\eta \eta} + G\phi^3,
\end{equation}
where $\mu > 0$.

Using Eq.~\eqref{eqn:5.102}, we can write down $\phi(\eta) = \sqrt{\mu}\tanh{\left[\sqrt{\frac{\mu}{2}}(\eta-\eta_0)\right]}$, the kink solution of Eq.~\eqref{eqn:1.4} with $G = 1$, as
\begin{equation}
\label{eqn:3.1}
\phi = \sqrt{\mu}\tanh{\left[\sqrt{\frac{\mu}{2}}\frac{3}{2} \left(\frac{e^{\frac{2}{3}x}}{c_1(t)} - \frac{e^{\frac{2}{3}x_0}}{c_1(t_0)}\right)\right]},
\end{equation}
where we have chosen $c_3 = \eta_0 -\frac{3}{2c_1(t_0)} e^{\frac{2}{3}x_0}$ to match with the soliton centre $x_0$ at time $t = t_0$.

Therefore, using Eqs.~\eqref{eqn:5.1}, \eqref{eqn:5.100}, \eqref{eqn:5.101} and~\eqref{eqn:3.1}, we have constructed the solution
\begin{equation}
\label{eqn:h1}
\begin{split}
\psi(R,T) &= \frac{\sqrt{\mu}}{c_1(T)R^{\frac{1}{3}}}e^{i\left(\frac{3 \dot{c}_1(T)}{8c_1(T)}R^2 + c_2(T)\right)} \\
&\times \tanh{\left[\frac{3}{2 \sqrt{2}}\sqrt{\mu}\left(\frac{R^{\frac{2}{3}}}{c_1(T)} - \frac{R_0^{\frac{2}{3}}}{c_1(T_0)}\right)\right]} 
\end{split}
\end{equation}
of the original radial GPE, Eq.~\eqref{eqn:2} ($\sigma = 1$), with the potential
\begin{equation}
\label{eqn:h1pot}
\begin{split}
V(R,T) &= \frac{1}{9 R^2} + \frac{-\mu}{c_1^2(T)R^{\frac{2}{3}}} \\
&- \frac{3}{16} \frac{\dot{c}_1^2(T) + 2c_1(T)\ddot{c}_1(T)}{c_1^2(T)}R^2 - \dot{c}_2(T).
\end{split}
\end{equation}

\begin{figure}[ht]
\includegraphics[angle=-90,width=0.45\textwidth]{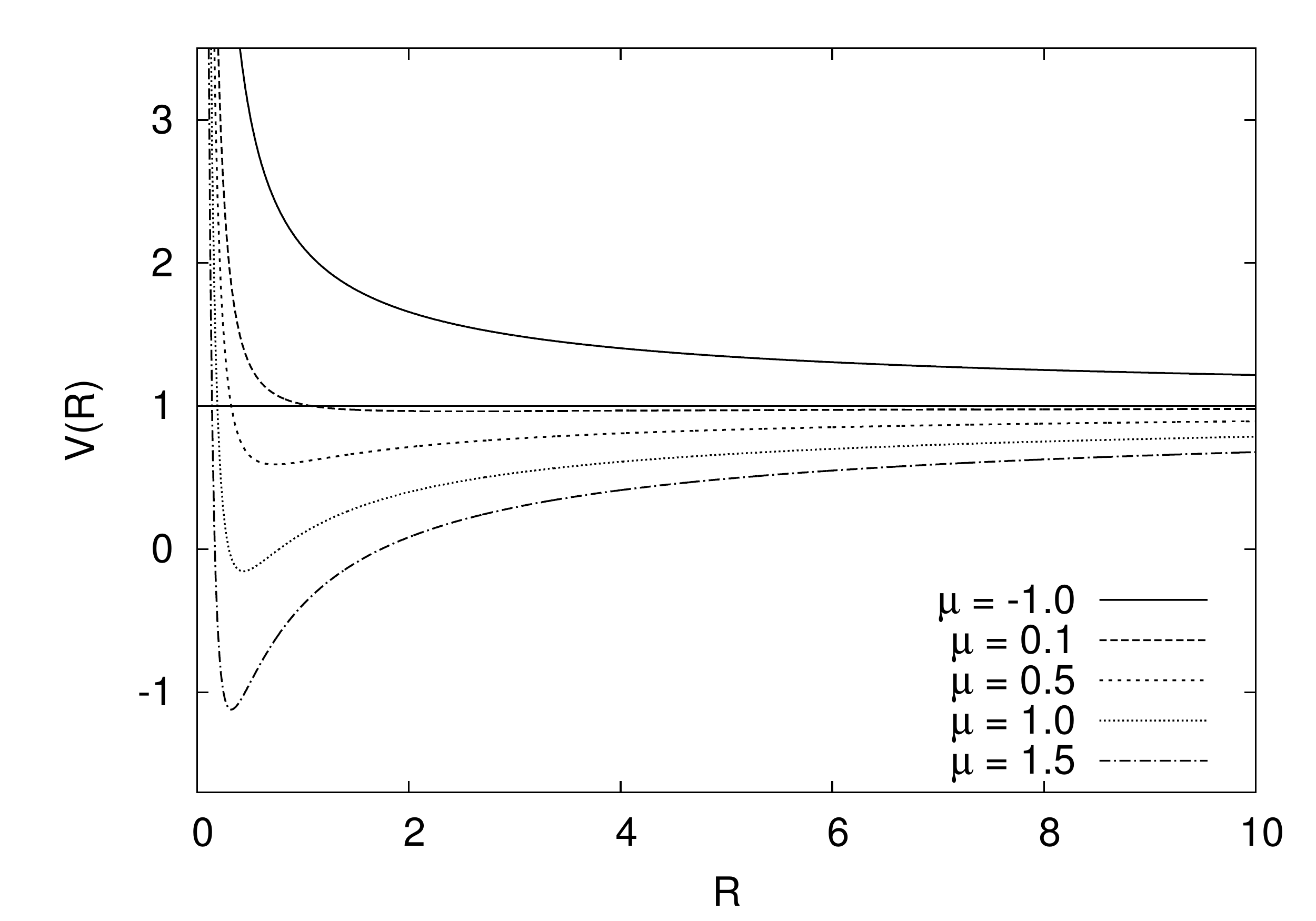}
\caption{\label{fig:pot} The potential~\eqref{eqn:3.5a} for a few values of $\mu$. There is a minimum at finite $R$ only for positive values of $\mu$. The potential approaches unity as $R \to \infty$.}
\end{figure}

\begin{figure}[ht]
  \includegraphics[angle=-90,width=0.45\textwidth]{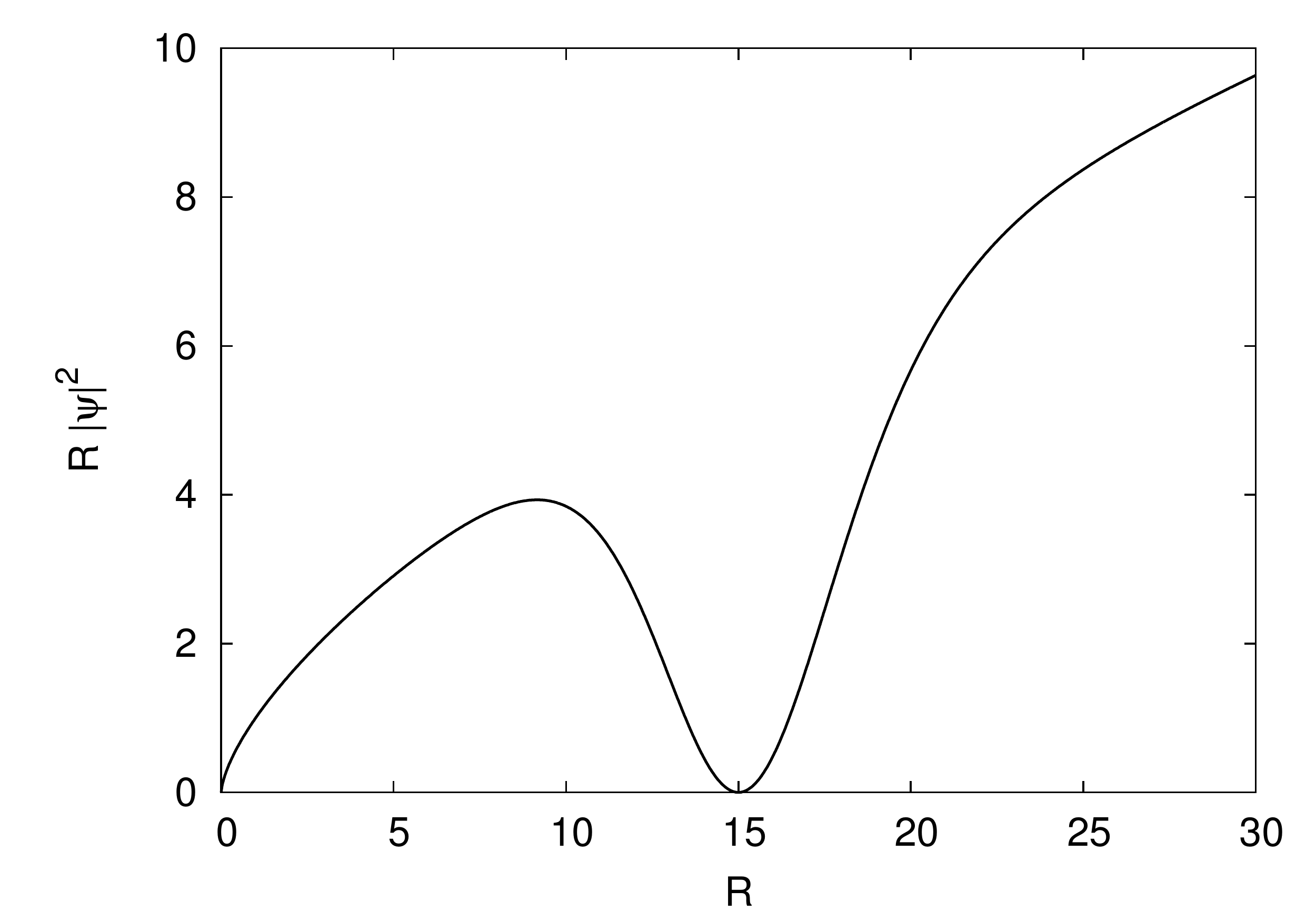}
  \caption{The ring dark soliton-like solution~\eqref{eqn:3.5} with $R_0 = 15$ and $\mu = 1$.}
  \label{fig:RDS}
\end{figure}

Let us select $c_1(T) = 1$ and $c_2(T) = -T$. Then from Eq.~\eqref{eqn:h1pot} we get (see Fig.~\ref{fig:pot})
\begin{equation}
\label{eqn:3.5a}
V(R) = \frac{1}{9R^2} - \mu \frac{1}{R^{\frac{2}{3}}} + 1 
\end{equation}
and from Eq.~\eqref{eqn:h1}
\begin{equation}
\label{eqn:3.5}
\psi(R,T) = R^{-\frac{1}{3}} \sqrt{\mu}\tanh{\left[\frac{3}{2 \sqrt{2}}\sqrt{\mu}(R^{\frac{2}{3}} - R_0^{\frac{2}{3}})\right]} e^{-iT},
\end{equation}
where $\mu > 0$ (see Fig.~\ref{fig:RDS}).

We note that it is possible to rewrite the $1/(9R^2)$ term in the potential as a fractional vorticity of $\frac{1}{3}$ and include the full Laplacian. Then
\begin{equation}
\label{eqn:3.5fracv}
\psi(R,T,\theta) = R^{-\frac{1}{3}} \sqrt{\mu}\tanh{\left[\frac{3}{2 \sqrt{2}}\sqrt{\mu}(R^{\frac{2}{3}} - R_0^{\frac{2}{3}})\right]} e^{-i(T-\frac{\theta}{3})}
\end{equation}
solves
\begin{equation}
\label{eqn:2fracv}
i\psi_{T} = -\psi_{RR} - \frac{1}{R} \psi_{R} - \frac{1}{R^2} \psi_{\theta \theta} + V(R) \psi + |\psi|^2 \psi,
\end{equation}
where 
\begin{equation}
\label{eqn:3.5afracv}
V(R) = - \mu \frac{1}{R^{\frac{2}{3}}} + 1 .
\end{equation}

\subsection{The stability of the dark solution, $h(\eta) = -\mu$}

Direct propagation of $\psi(x,y,t)$ given by a two-dimensional extension of the GPE equation~\eqref{eqn:nls} in rectangular coordinates $x,y$ shows that the solution given in Eq.~\eqref{eqn:3.5} is long-lived, until it decays into a vortex-antivortex necklace by the snake instability. The decay is much faster if instead of the potential in Eq.~\eqref{eqn:3.5a} we use e.g.
\begin{equation}
\label{eqn:3.5aa}
\tilde{V}(R) = \frac{A}{R^3} - \frac{B}{R},
\end{equation}
where $A$ and $B$ are chosen so that $V(R)$ and $\tilde{V}(R)$ agree up to the first (linear) order when expanded at $R = R_0$.

\begin{figure}
  \centering
  \includegraphics[angle=-90,width=0.45\textwidth]{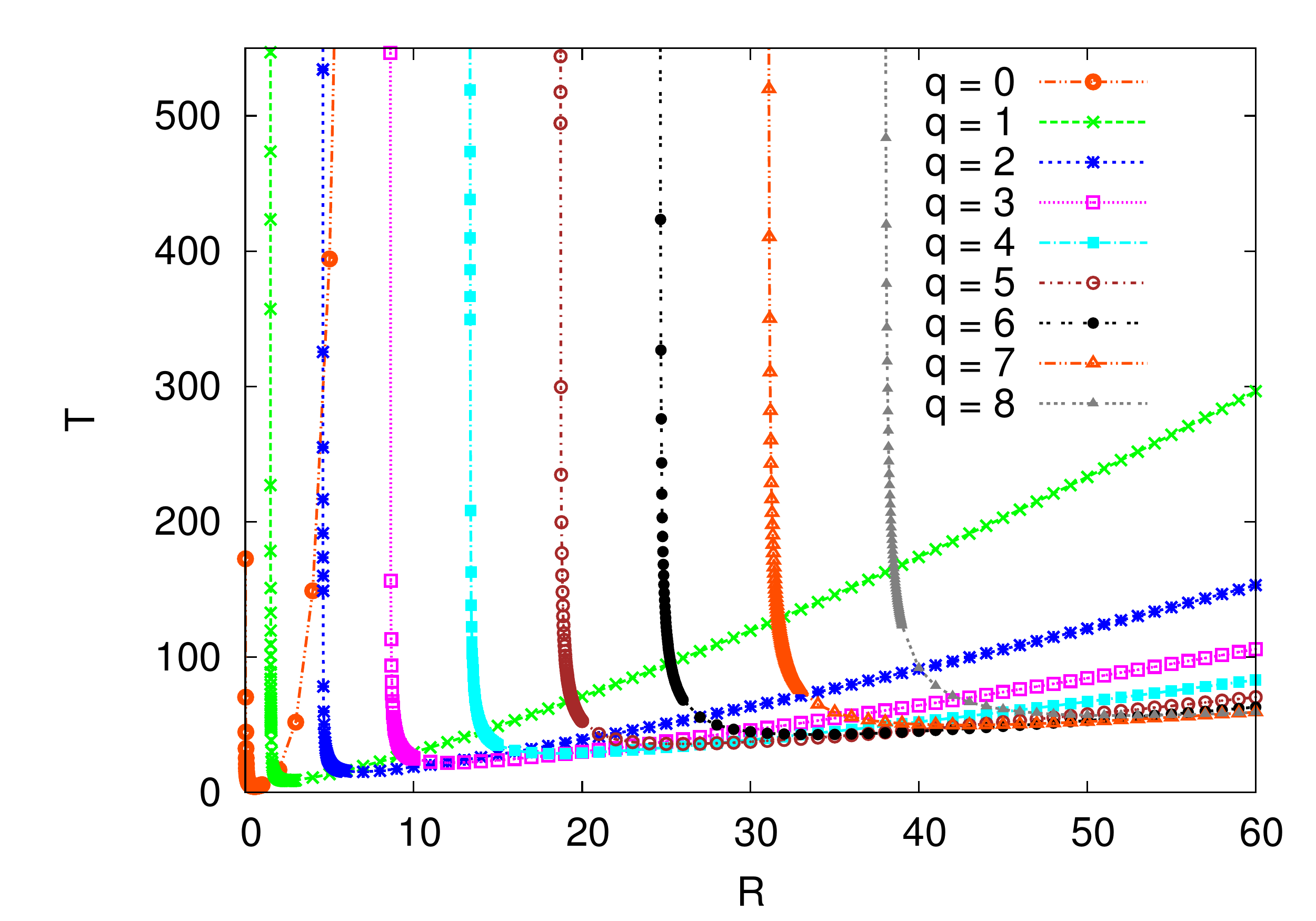}
  \caption{(Colour online.) Results of solving the Bogoliubov equations~\eqref{BdG_matrix_natunits} for the exact solution~\eqref{eqn:3.5} of Eq.~\eqref{eqn:2} ($\sigma = 1$) with the potential~\eqref{eqn:3.5a}. Here $\lambda = 1$. Shown are the decay times $1 / \mathrm{Im}(\epsilon)$, where $\epsilon$ is the Bogoliubov eigenvalue. There is only a single imaginary eigenvalue per $q$ and the corresponding amplitude is localised at the notch of the ring dark soliton-like solution. For low radii ($R_S \lesssim 1.5003$) the only unstable mode is $q = 0$. For radii $4.613 \lesssim R_S \lesssim 8.620$ the primary decay channel is the quadrupole, or $q = 2$, mode, while all higher modes are dynamically stable. As the radius is increased the higher modes become unstable as well. The solution~\eqref{eqn:3.5} is always formally unstable with respect to the $q = 0$ mode, although the instability times are very long for large radii.}
  \label{fig:BdGE_analytical}
\end{figure}

To get more insight on the stability we consider the Bogoliubov-de-Gennes equations for low-lying modes $\epsilon$~\cite{Dauxois2006,Kinks2006,P&S}. In dimensionless natural units ($\mu = \hbar = 2m = 1$) they are
\begin{equation}
\label{BdG_matrix_natunits}
\begin{pmatrix} \mathcal{L} & \psi_0^2 \\ \bar{\psi}_0^2 & \mathcal{L} \end{pmatrix} 
\begin{pmatrix} u_q \\ v_q \end{pmatrix} = \epsilon \begin{pmatrix} u_q \\ -v_q \end{pmatrix},
\end{equation}
where we have defined the Bogoliubov amplitudes $u$ and $v$ for which
\begin{equation}
\label{BogExp}
\psi(\textbf{r},T) =  e^{-i T}\lbrace \psi_0(\textbf{r}) + [u(\textbf{r})e^{-i\epsilon T} + \bar{v}(\textbf{r})e^{i\epsilon T}] \rbrace,
\end{equation}
by~\cite{PhysRevA.58.3168}
\begin{equation}
\label{eqn:BdGEpolarcoords}
\begin{pmatrix} u(\textbf{r}) \\ v(\textbf{r}) \end{pmatrix} = e^{iq\theta} \begin{pmatrix} u_q(R) \\ v_q(R) \end{pmatrix},
\end{equation}
representing a partial wave of angular momentum $q$ relative to the condensate, and
\begin{equation}
\mathcal{L} \equiv -\left(\frac{\partial^2}{\partial R^2} + \frac{1}{R}\frac{\partial}{\partial R} - \frac{q^2}{R^2}\right)  + \frac{1}{9R^2} - \frac{\lambda}{R^{\frac{2}{3}}} + 2|\psi_0|^2,
\end{equation}
and $\psi_0(R,T)$ is given by Eq.~\eqref{eqn:3.5} (with the substitution $\mu \rightarrow \lambda$). If $\epsilon > 0$, $u$ and $v$ are orthogonal and normalised according to
\begin{equation}
\label{eqn:BdGEuvnorm}
\int d\textbf{r} (|u(\textbf{r})|^2- |v(\textbf{r})|^2) = 1.
\end{equation}
The instability time is given by $T_d = 1 / \mathrm{Im}(\epsilon)$, and if all the eigenvalues are real, the condensate is dynamically stable. The results of numerically solving Eq.~\eqref{BdG_matrix_natunits} are shown in Fig.~\ref{fig:BdGE_analytical}. There is only one (purely) imaginary eigenvalue, and the Bogoliubov amplitudes are localised at the radius $R_S$. Equation~\eqref{BdG_matrix_natunits} is solved numerically using Lagrange functions, choosing a Laguerre discrete-variable representation \cite{0305-4470-19-11-013, DBLP:journals/cphysics/McPeakeM04}, which is particularly efficient for radial geometry. The abscissas can be found using Newton's method \cite{NR}, and the resulting linear eigenvalue problem is solved by Hessenberg QR iteration \cite{laug}.

The minimum of the potential~\eqref{eqn:3.5a} with $\lambda = 1$ occurs at $R_0 = \left( \frac{1}{3} \right)^{\frac{3}{4}} \approx 0.44$. Therefore, the soliton-like solution~\eqref{eqn:3.5} is stable if $R_S$ is around (a healing length away from) the minimum of the potential. Another region of stability appears to be the limit $R_S \to \infty$ (see Fig.~\ref{fig:BdGE_analytical}), although it is approached rather slowly. In this limit we obtain the one-dimensional Gross-Pitaevskii equation with a constant potential, which can be removed by redefining the zero of energy.
\begin{figure}
  \centering
  \includegraphics[angle=-90,width=0.45\textwidth]{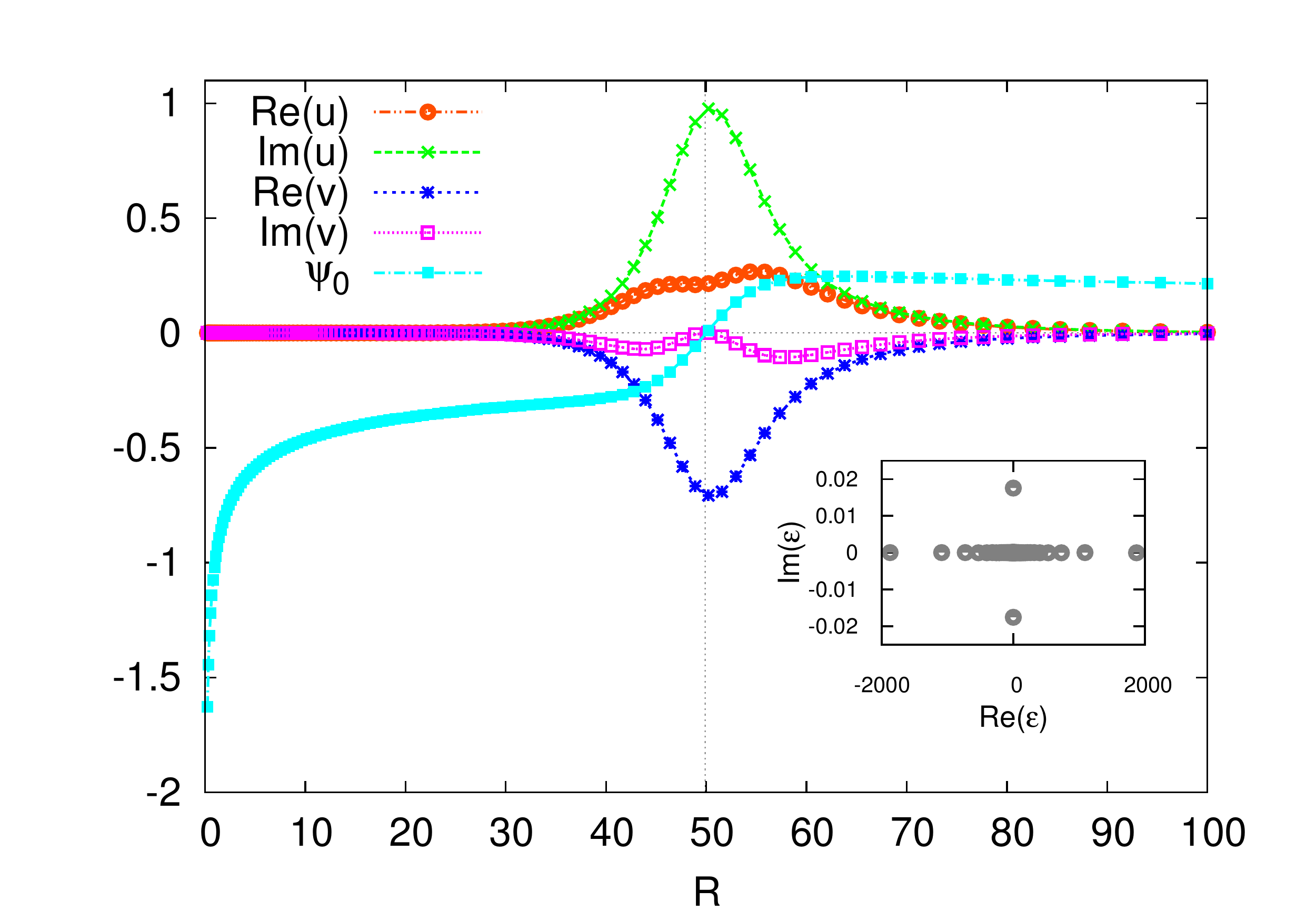}
  \caption{(Colour online.) The amplitudes $u$ and $v$ corresponding to the only imaginary Bogoliubov mode, and the wavefunction $\psi_0$ for $R_S = 50.0$, $\lambda = 1$, $q = 8$, and $\theta = 0$ at $T = 0$. The inset shows the Bogoliubov spectrum $\epsilon$. The amplitudes $u$ and $v$ are localised around the notch at $R_S = 50.0$. We have chosen one possible normalisation satisfying~\eqref{eqn:BdGEuvnorm}.}
  \label{fig:BdGEuv}
\end{figure}

\begin{figure}
  \centering
  \subfigure[$\theta = 0$]{
  \includegraphics[angle=-90,width=0.45\textwidth]{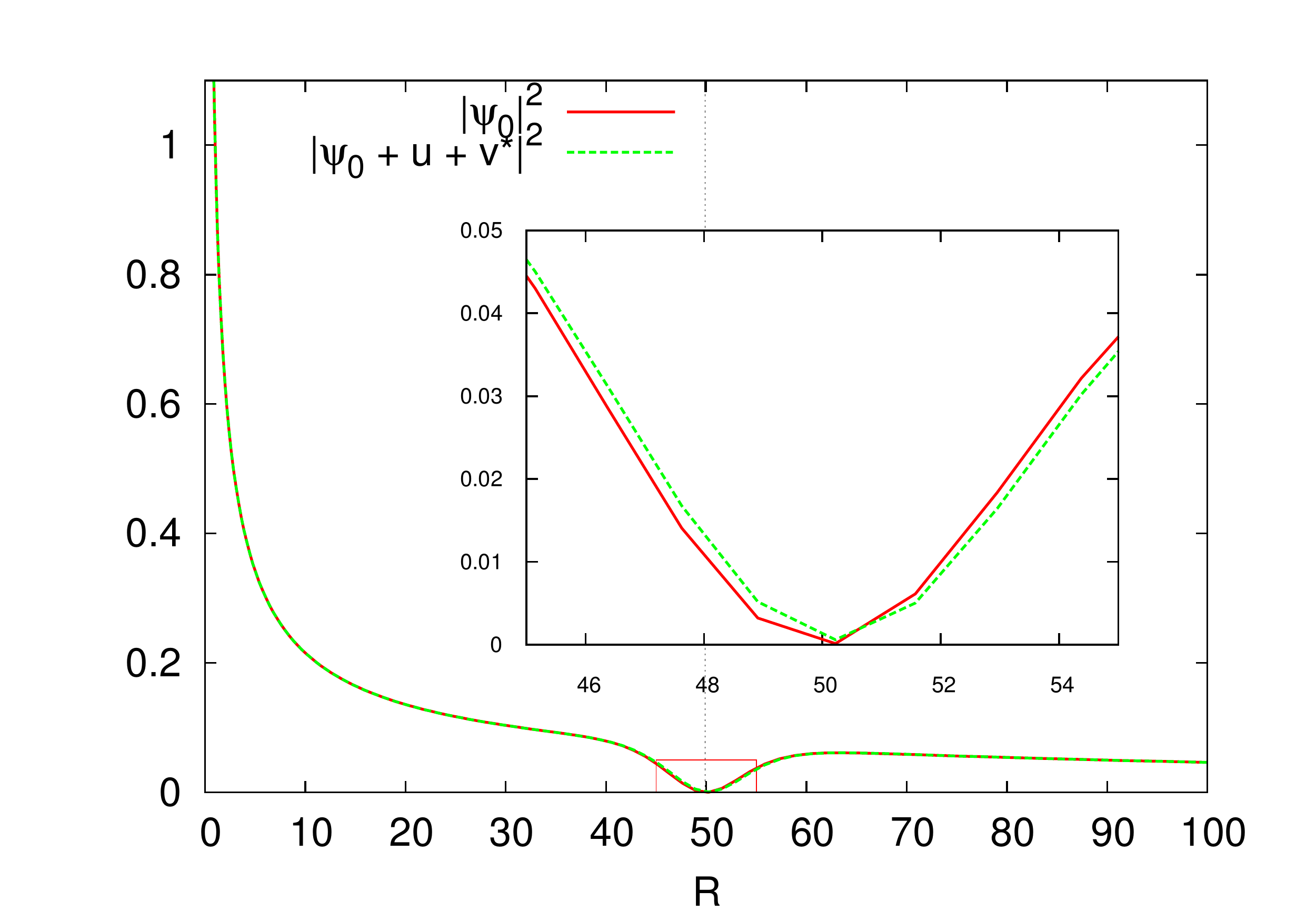}
	\label{subfig:BdGEwfuva}
  }
  \subfigure[$\theta = \pi/8$]{
  \includegraphics[angle=-90,width=0.22\textwidth]{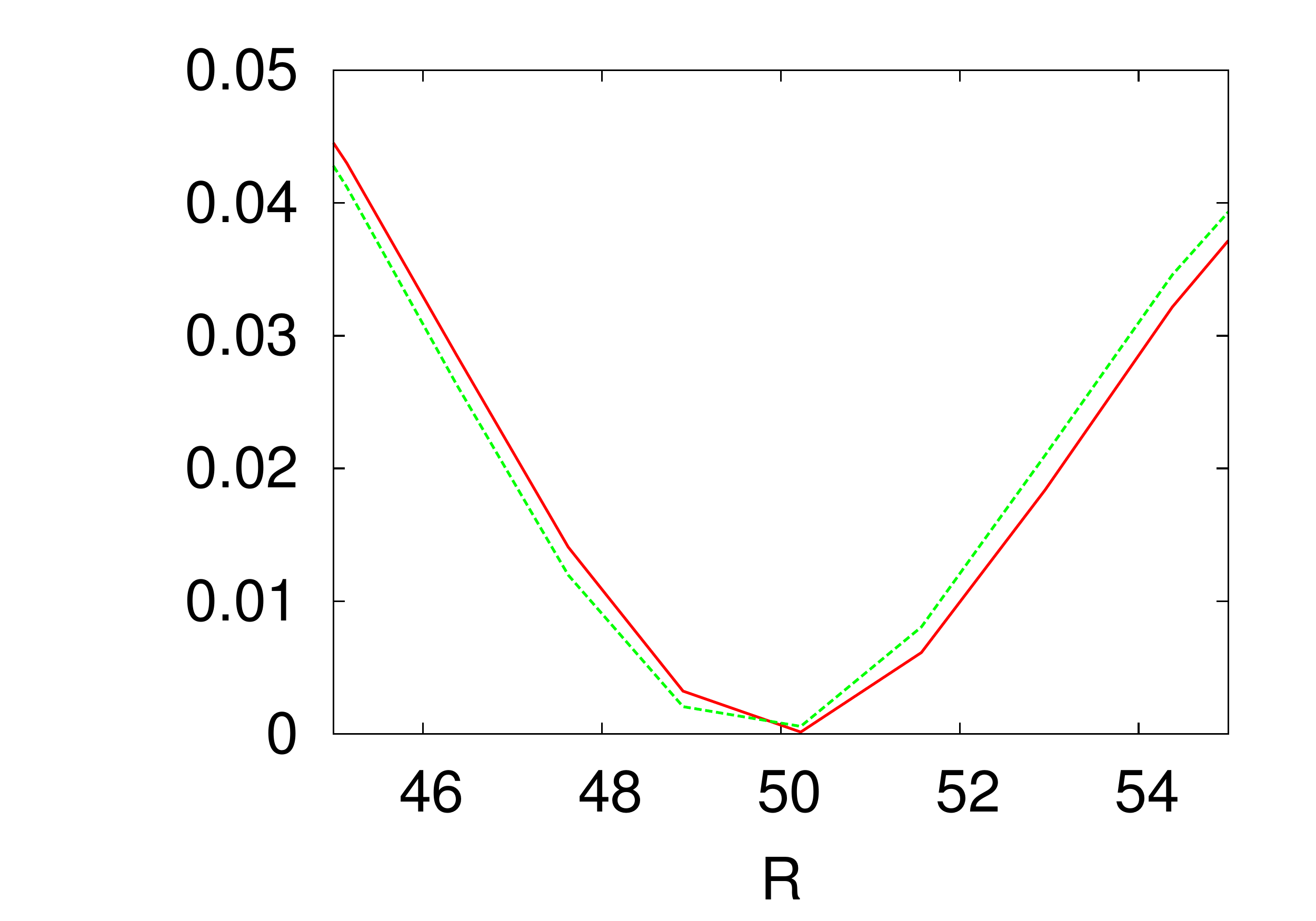}
	\label{subfig:BdGEwfuvb}
  }  
  \subfigure[$\theta = 2\pi/8$]{
  \includegraphics[angle=-90,width=0.22\textwidth]{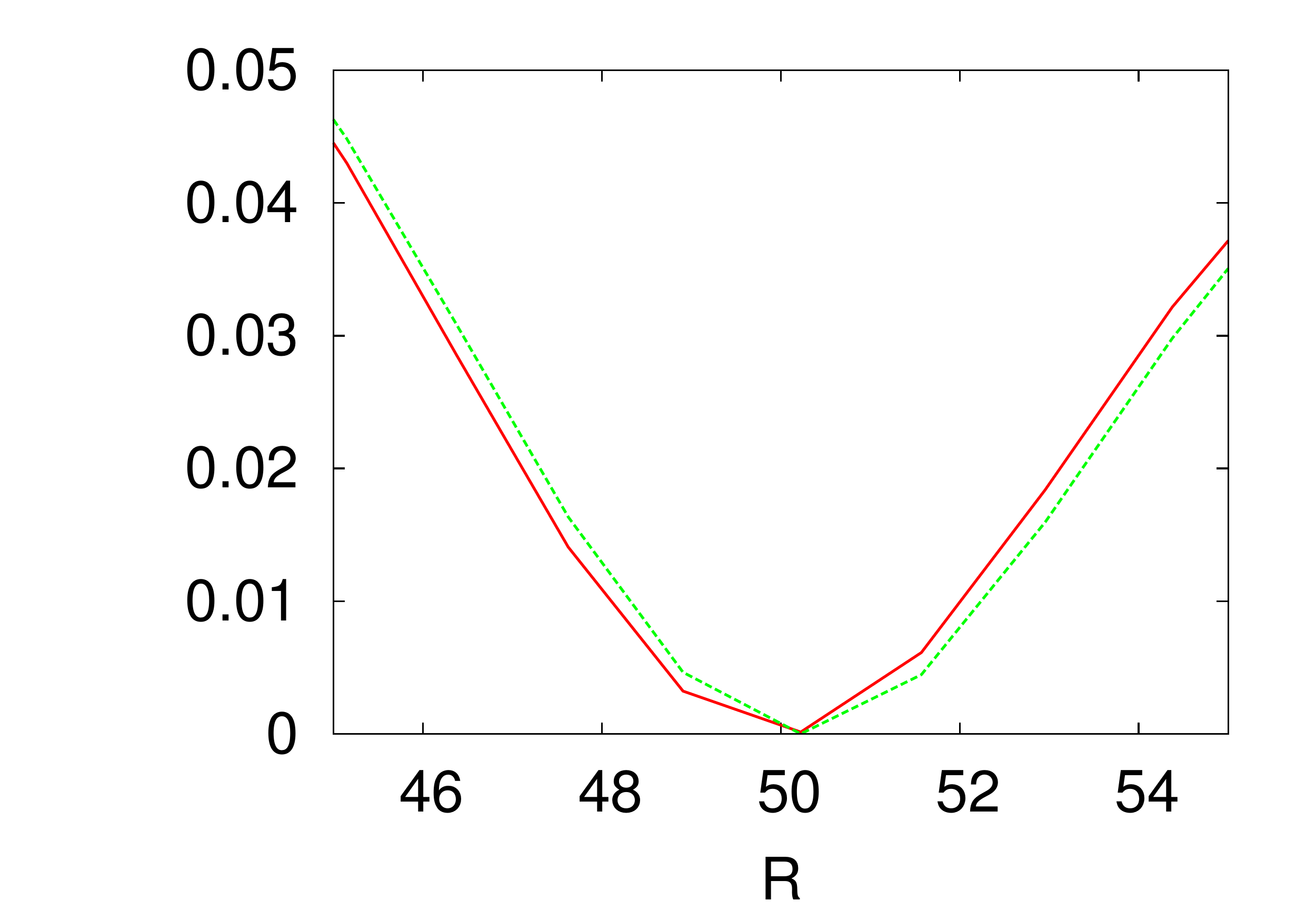}
	\label{subfig:BdGEwfuvc}
  }  
  \subfigure[$\theta = 3\pi/16$]{
  \includegraphics[angle=-90,width=0.22\textwidth]{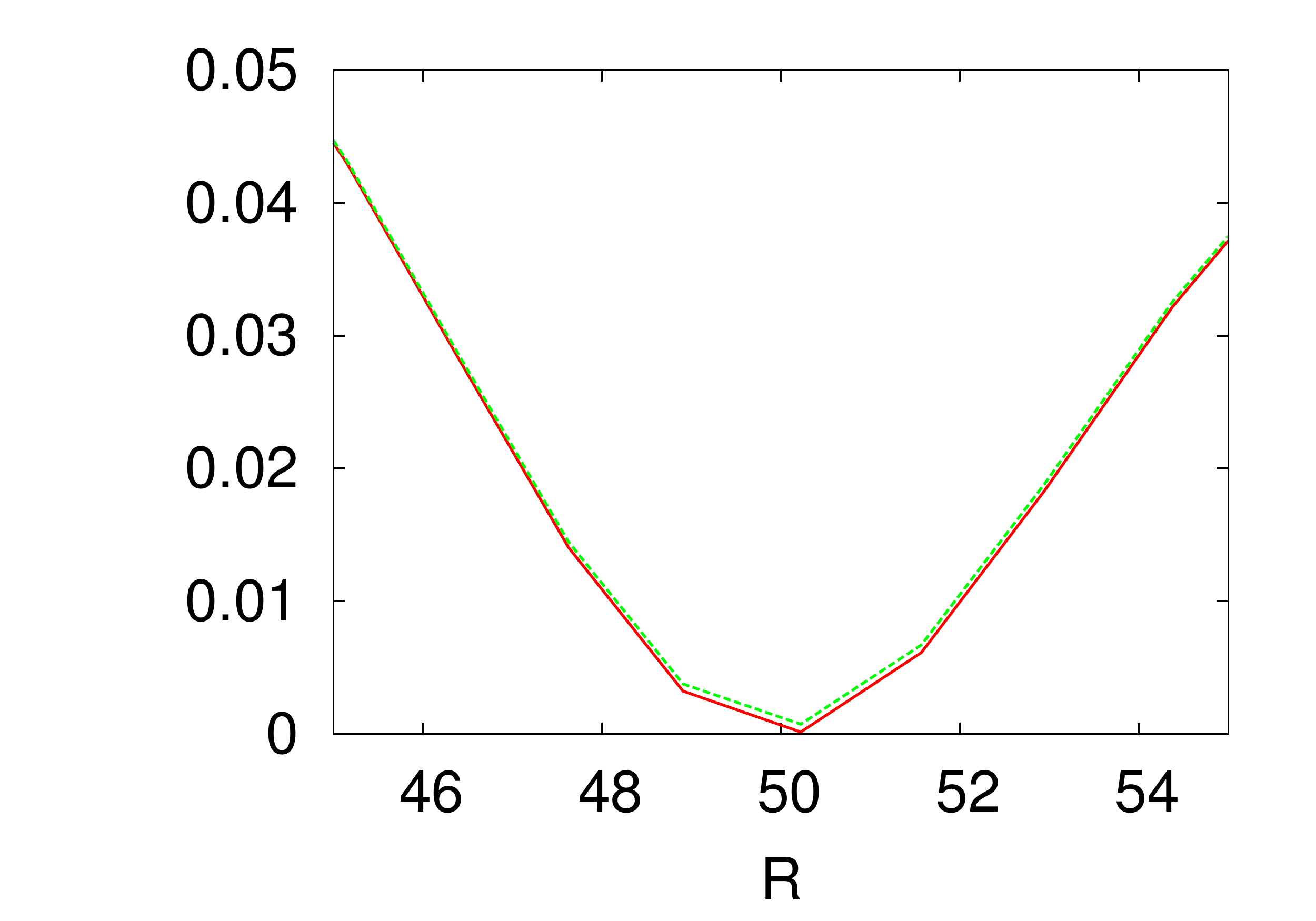}
	\label{subfig:BdGEwfuvd}
  }
  \subfigure[$\theta = 7\pi/32$]{
  \includegraphics[angle=-90,width=0.22\textwidth]{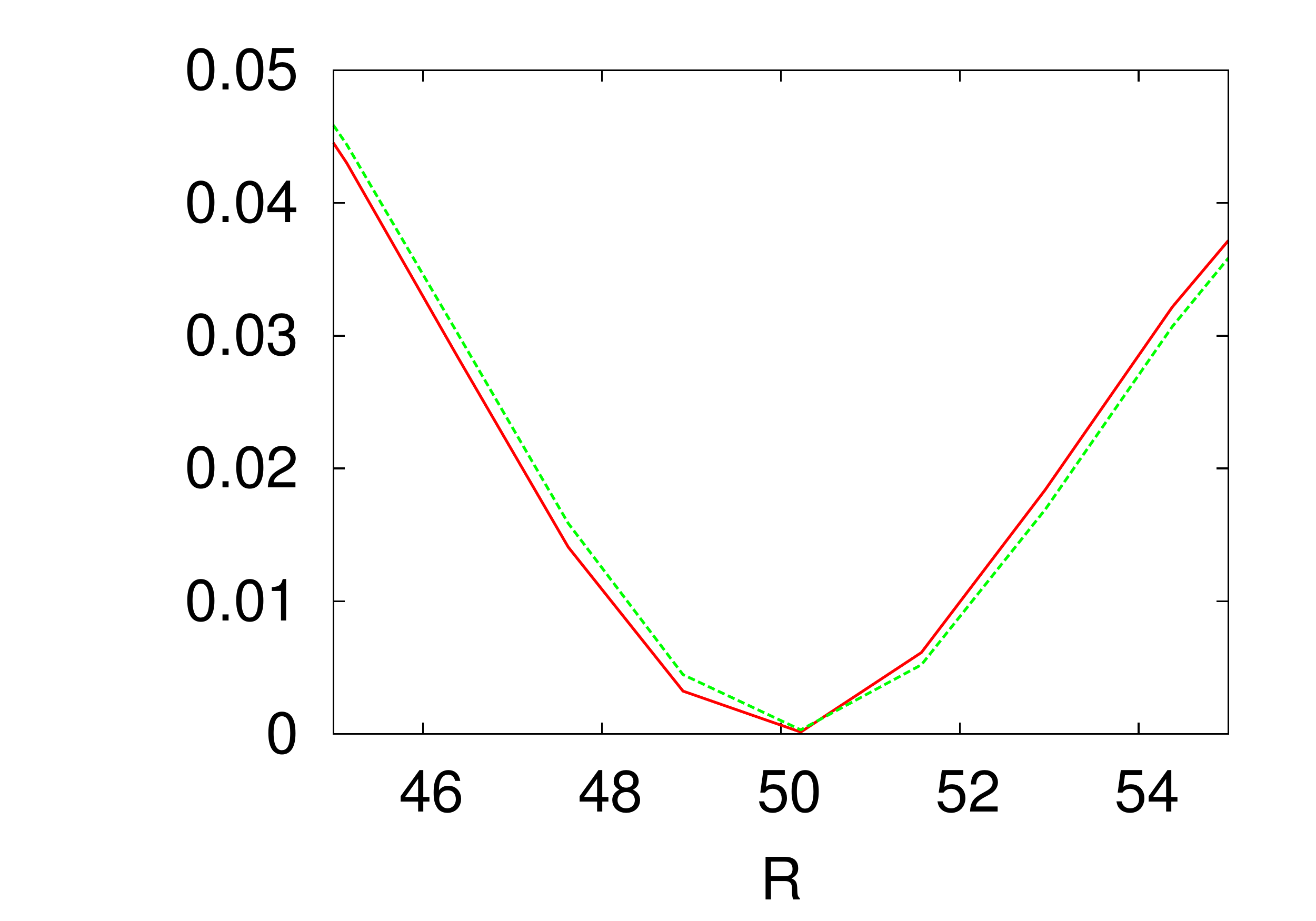}
	\label{subfig:BdGEwfuve}
  }   
  \caption{(Colour online.) The instability mode is directly related to the snake instability, having typical behaviour around the ring (for $q = 8$). \subref{subfig:BdGEwfuva} The amplitudes $u$ and $v$ of Fig.~\ref{fig:BdGEuv} added to the wavefunction $\psi_0$ at $\theta = 0$. The disturbance is towards increasing radius. The inset shows a magnification of the area inside the red rectangle. \subref{subfig:BdGEwfuvb} At $\theta = \pi/8$, the disturbance is now towards decreasing radius. \subref{subfig:BdGEwfuvc} At $\theta = 2\pi/8$, the disturbance is again towards increasing radius. \subref{subfig:BdGEwfuvd} At $\theta = 3\pi/16$, between the previous extremes, the disturbance does nothing. \subref{subfig:BdGEwfuve} At $\theta = 7\pi/32$, which is not an extremal angle, the disturbance is also between extremes.}
  \label{fig:BdGEwfuv}
\end{figure}
The case for $R_S = 50.0$, $\lambda = 1$, and $q = 8$ is shown in more detail in Figs.~\ref{fig:BdGEuv} and~\ref{fig:BdGEwfuv}. Because the ring suffers from the snake instability in general, and because there is only one mode with dynamical instability (per $q$), they must be related. That this is indeed so is confirmed in Figs.~\ref{fig:BdGEuv} and~\ref{fig:BdGEwfuv}. The amplitudes $u$ and $v$ corresponding to the imaginary eigenvalue are localised at the notch. Here the azimuthal symmetry was broken by choosing~\eqref{eqn:BdGEpolarcoords}, that is, an overall phase of 0.

\subsection{The bright ring soliton-like solution, $h(\eta) = -\mu$}
\label{sec:RBS}

\begin{figure}[ht]
  \centering
  \includegraphics[angle=-90,width=0.45\textwidth]{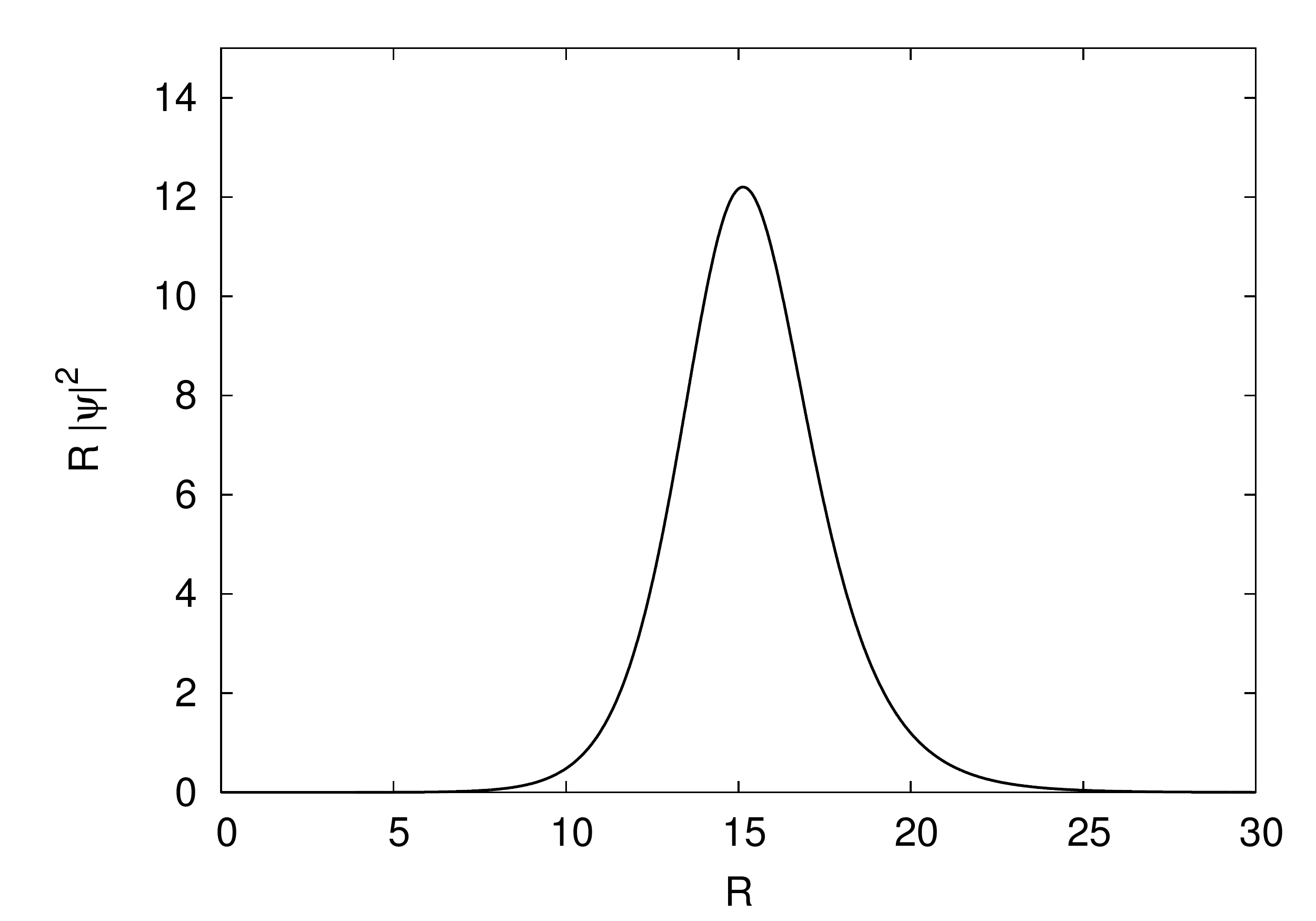}
  \caption{The ring bright soliton-like solution~\eqref{eqn:3.6} with $R_0 = 15$ and $\mu = -1$.}
  \label{fig:RBS}
\end{figure}

When $G = -1$, Eq.~\eqref{eqn:1.4} has also a bright soliton-like solution $\phi(\eta) = \sqrt{2}\sqrt{-\mu} \sech{[\sqrt{-\mu}(\eta-\eta_0)]}$ ($\mu < 0$). Similarly as in Sec.~\ref{sec:kink} (now $\alpha_3 = G = -1$ so nothing changes), we arrive at the solutions
\begin{equation}
\label{eqn:h1b}
\begin{split}
\psi(R,T) &= \frac{\sqrt{-2\mu}}{c_1(T)R^{\frac{1}{3}}}e^{i\left(\frac{3 \dot{c}_1(T)}{8c_1(T)}R^2 + c_2(T)\right)} \\
&\times \sech{\left[\frac{3}{2}\sqrt{-\mu}\left(\frac{R^{\frac{2}{3}}}{c_1(T)} - \frac{R_0^{\frac{2}{3}}}{c_1(T_0)}\right)\right]} 
\end{split}
\end{equation}
with the potential~\eqref{eqn:h1pot} ($\mu < 0$) and
\begin{equation}
\label{eqn:3.6}
\psi(R,T) = R^{-\frac{1}{3}} \sqrt{-2 \mu}\sech{\left[\frac{3}{2}\sqrt{-\mu}(R^{\frac{2}{3}} - R_0^{\frac{2}{3}})\right]} e^{-iT}
\end{equation}
with the potential~\eqref{eqn:3.5a}, where $\mu < 0$ (see Fig.~\ref{fig:RBS}), of the original radial GPE, Eq.~\eqref{eqn:2} ($\sigma = -1$).

\section{Painlev\'e II, $h(\eta) = \eta$}\label{sec:p22}

We are now requiring $\phi(\eta)$ to satisfy
\begin{equation}
\label{eqn:12.0}
-\phi_{\eta \eta} + \eta \phi + 2\phi^3 = 0.
\end{equation}
This means $g(\eta) = 0, h(\eta) = \eta,$ and $G = 2$. As we can see in Table~\ref{tab:1}, Eq.~\eqref{eqn:12.0} arises as the similarity reduction of Eq.~\eqref{eqn:2} ($\sigma = 1$) with 
\begin{equation}
\label{eqn:12.1}
\begin{split}
V(R, T) &= \frac{1}{9 R^2} - \frac{3}{16} \frac{\dot{c}_1^2(T) + 2c_1(T)\ddot{c}_1(T)}{c_1^2(T)}R^2 \\
&+ \frac{3}{2^{\frac{5}{2}} c_1^3(T)} - \dot{c}_2(T).
\end{split}
\end{equation}
This is similar to what happens when the modified KdV equation
\begin{equation}
\label{eqn:12.2}
v_t - 6v^2v_x + v_{xxx} = 0
\end{equation}
is transformed with a similarity reduction 
\begin{equation}
\label{eqn:12.3}
v(x,t) = \frac{w(z)}{(3t)^{\frac{1}{3}}}, \qquad z = \frac{x}{(3t)^{\frac{1}{3}}}
\end{equation}
such that $w(z)$ satisfies $\mathrm{P_{II}}$:
\begin{equation}
\label{eqn:12.4}
w_{zz} = 2w^3 + zw + \alpha,
\end{equation}
where $\alpha$ is a constant of integration~\cite{PhysRevLett.38.1103}.

Interestingly, we can eliminate time dependence from the potential~\eqref{eqn:12.1} without sacrificing nontrivial time dependence in the solution, unlike what happened in Sections~\ref{sec:kink} and~\ref{sec:RBS}, by choosing
\begin{align}
\label{eqn:12.6}
c_1(T) &= c_4 \cosh{[2 \sqrt{k}(T + c_5)]^{\frac{2}{3}}}, \\
\label{eqn:12.5}
\dot{c}_2(T) &= \frac{3}{2^{\frac{5}{2}} c_1^3(T)} - c_6,
\end{align}
where $k \neq 0$, $c_4$, $c_5$, and $c_6$ are constants. 

Therefore,
\begin{equation}
\label{eqn:12.8}
\psi(R,T) = \rho(R,T)e^{i \varphi(R,T)} \phi(\eta(R,T)),
\end{equation}
where
\begin{align}
\label{eqn:12.8a}
\rho(R,T) &= \frac{R^{-\frac{1}{3}}}{c_4 \cosh{[2 \sqrt{k}(T + c_5)]^{\frac{2}{3}}}},\\
\label{eqn:12.8b}
\varphi(R,T) &= \frac{\sqrt{k}}{2}\tanh{[2\sqrt{k}(T+c_5)]} R^{2} + c_7 \notag \\
&- c_6(T + c_5) + \frac{3\tanh{[2\sqrt{k}(T+c_5)]} }{\sqrt{k}2^{\frac{7}{2}}c_4^3}, \\
\label{eqn:12.8c}
\eta(R,T) &= \frac{3}{2^{\frac{3}{2}}c_4 \cosh{[2 \sqrt{k}(T + c_5)]^{\frac{2}{3}}}} R^{\frac{2}{3}},
\end{align}
and where $c_7$ is a constant, solves Eq.~\eqref{eqn:2} ($\sigma = 1$) with the potential~\eqref{eqn:12.1}
\begin{equation}
\label{eqn:12.7}
V(R) = \frac{1}{9 R^2} - kR^2 + c_6,
\end{equation}
if $\phi(\eta)$ solves Eq.~\eqref{eqn:12.0}, the second Painlev\'e equation. 

If $k < 0$, we obtain from Eq.~\eqref{eqn:12.6}
\begin{equation}
\label{eqn:12.1001}
c_1(T) = c_4 \cos{[2 \sqrt{|k|}(T + c_5)]^{\frac{2}{3}}}.
\end{equation}
Then Eq.~\eqref{eqn:12.8} with 
\begin{align}
\label{eqn:12.1008a}
\rho(R,T) &= \frac{R^{-\frac{1}{3}}}{c_4 \cos{[2 \sqrt{|k|}(T + c_5)]^{\frac{2}{3}}}},\\
\label{eqn:12.1008b}
\varphi(R,T) &= -\frac{\sqrt{|k|}}{2}\tan{[2\sqrt{|k|}(T+c_5)]} R^{2} + c_7 \notag \\
&- c_6(T + c_5) + \frac{3\tan{[2\sqrt{|k|}(T+c_5)]} }{\sqrt{|k|}2^{\frac{7}{2}}c_4^3}, \\
\label{eqn:12.1008c}
\eta(R,T) &= \frac{3}{2^{\frac{3}{2}}c_4 \cos{[2 \sqrt{|k|}(T + c_5)]^{\frac{2}{3}}}} R^{\frac{2}{3}}
\end{align}
solves Eq.~\eqref{eqn:2} with the binding trap potential
\begin{equation}
\label{eqn:12.1007}
V(R) = \frac{1}{9 R^2} + |k|R^2 + c_6.
\end{equation}

\begin{figure*}[ht]
\centering
\subfigure[$k = -0.1$ and $c_5 = \frac{\pi}{4\sqrt{0.1}}$]{
   \includegraphics[angle=-90,width=0.2985\textwidth] {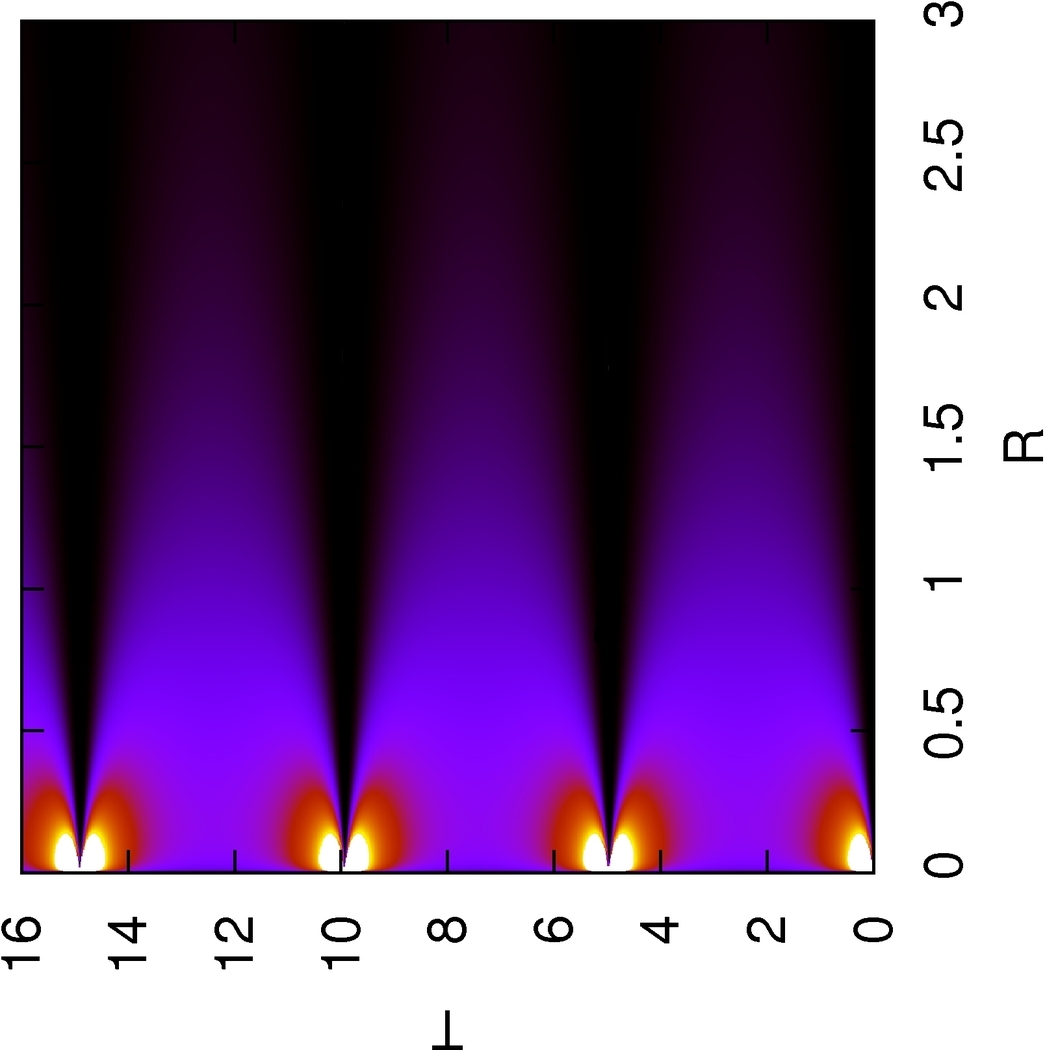}   
\label{fig:subfig4.1a}
 }
\subfigure[$k = -0.02$ and $c_5 = \frac{\pi}{4\sqrt{0.02}}$]{
   \includegraphics[angle=-90,width=0.2985\textwidth] {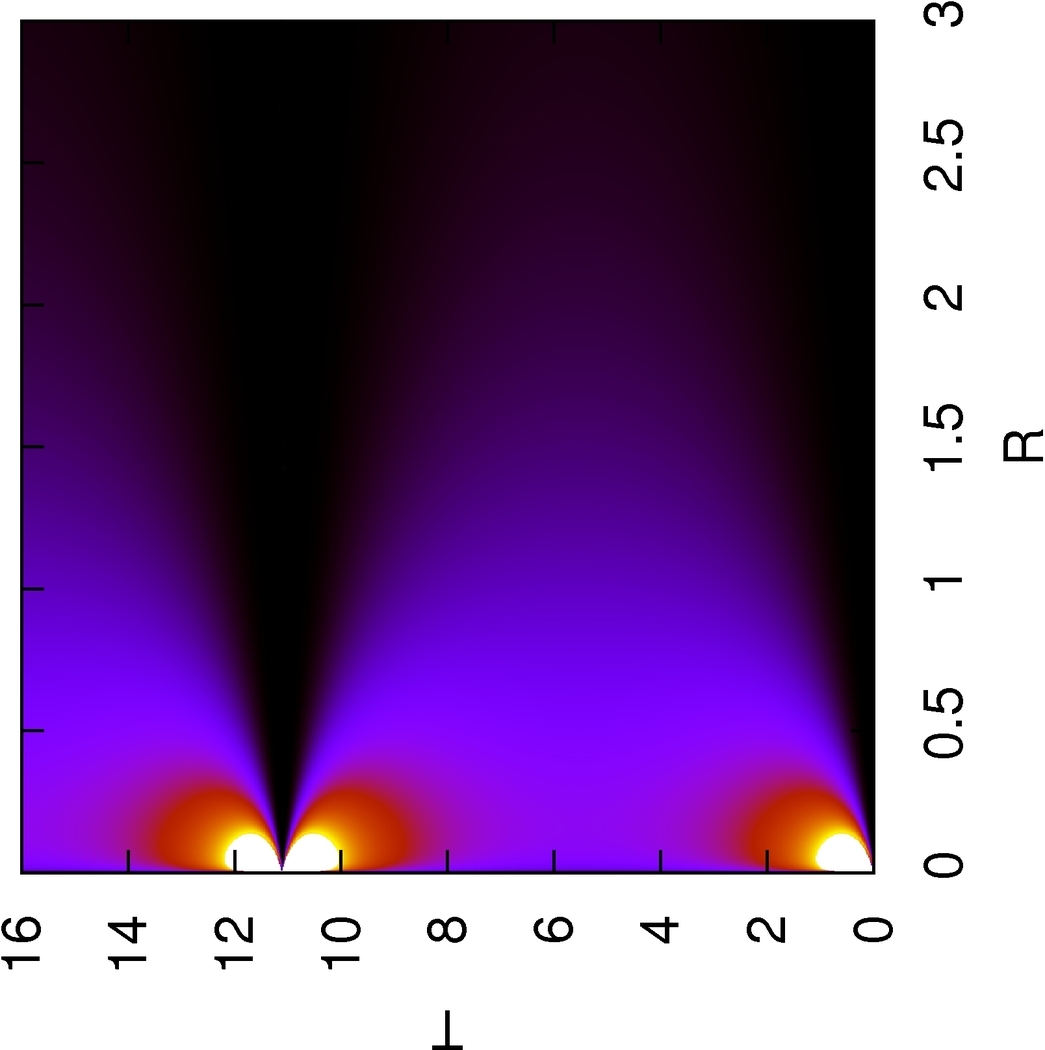}   
   \label{fig:subfig4.1b}
 }
\subfigure[$k = 2.0$ and $c_5 = 0$]{
  	\includegraphics[angle=-90,width=0.303\textwidth] {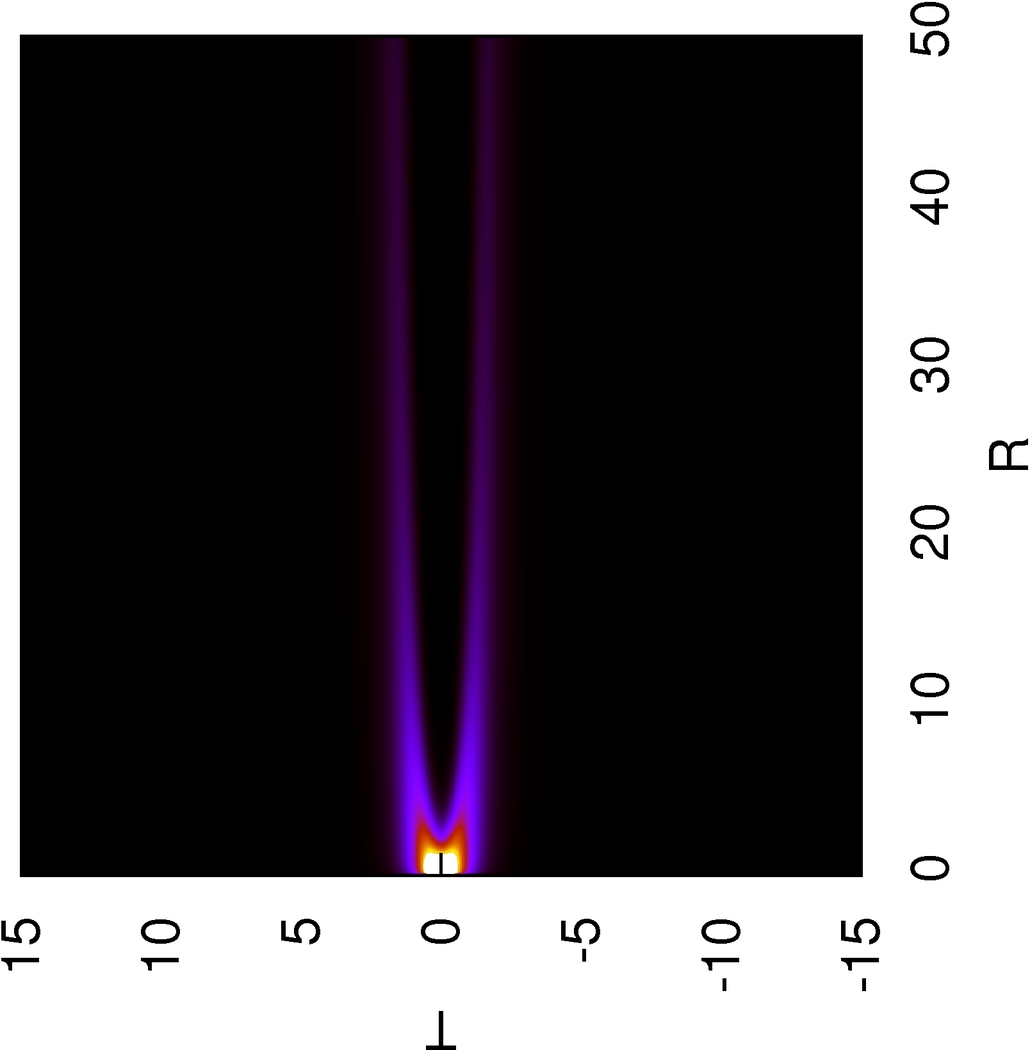}   
   \label{fig:subfig4.1c}
 }   
\subfigure[$k = 0.01$ and $c_5 = 0$]{
   \includegraphics[angle=-90,width=0.2805\textwidth] {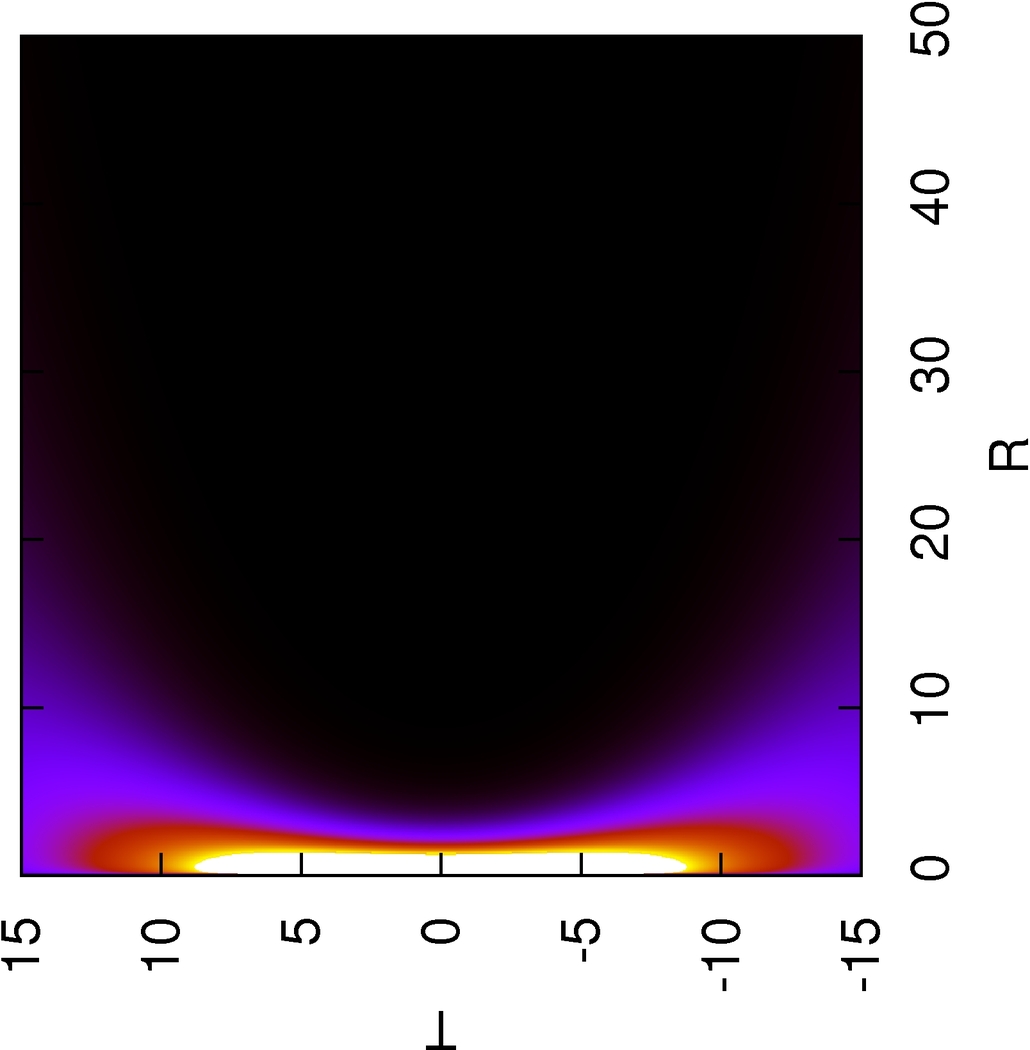}   
   \label{fig:subfig4.1d}
 } 
\subfigure[$k = 0.001$ and $c_5 = 0$]{
   \includegraphics[angle=-90,width=0.3\textwidth] {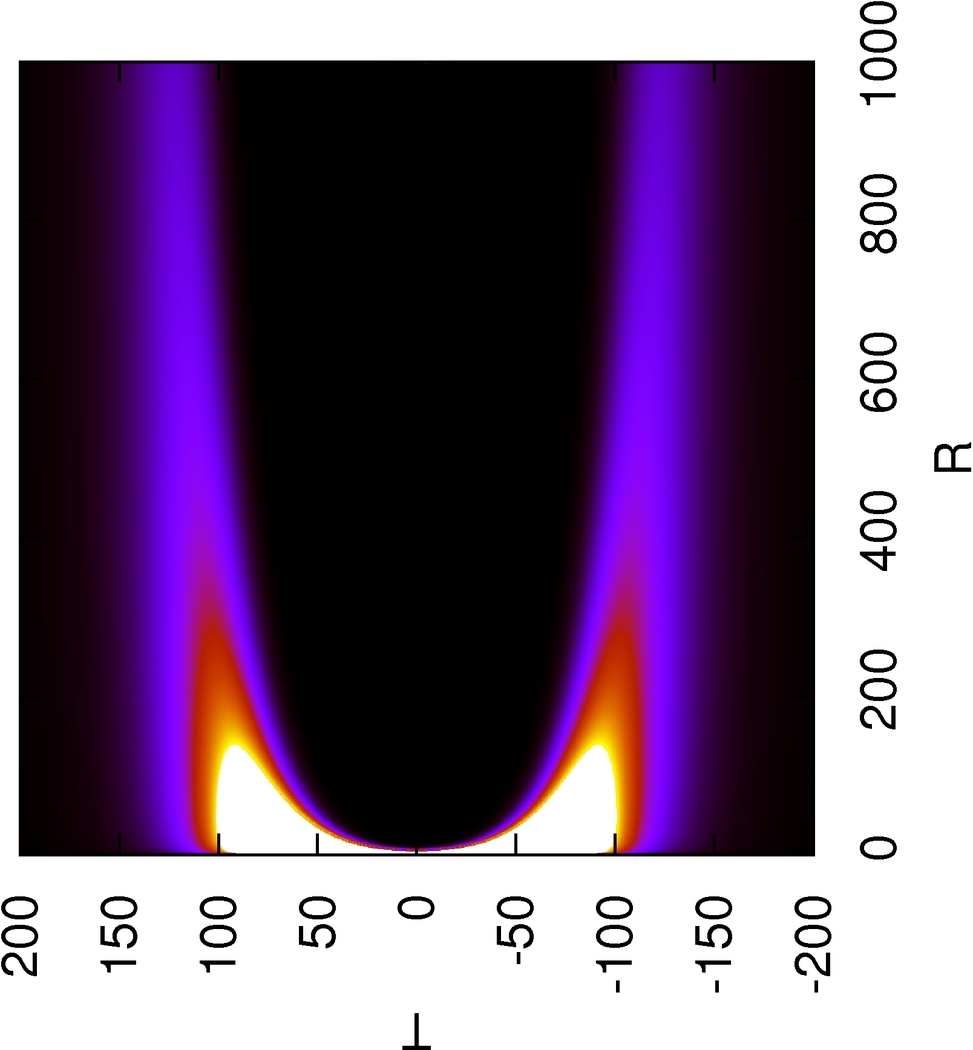}   
   \label{fig:subfig4.1e}
 }  
\subfigure[$k = 0$ and $c_5 = 0$]{
   \includegraphics[angle=-90,width=0.3105\textwidth] {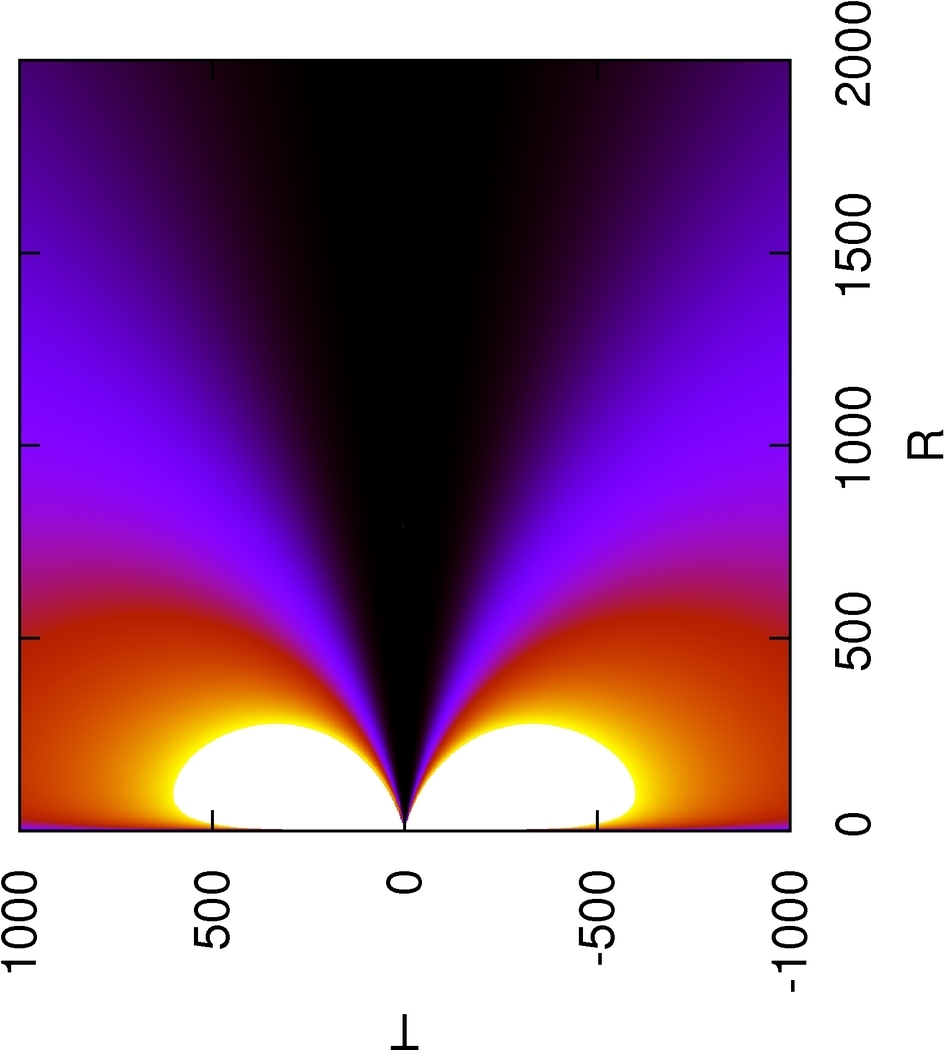}   
   \label{fig:subfig4.1f}
 }  
\caption{\label{Fig:4.1}(Colour online.) Density plots of $R|\psi|^2$ as given by Eq.~\eqref{eqn:12.8}. In all of the plots $\nu = 0.9$ and $c_4 = 3/2^\frac{3}{2}$. In~\subref{fig:subfig4.1a} and~\subref{fig:subfig4.1b} the radius of the centre of mass is oscillating and the condensate is breathing between a point and an extended state. The magnitude of $k$ determines the period of the oscillations. In~\subref{fig:subfig4.1c},~\subref{fig:subfig4.1d}, and~\subref{fig:subfig4.1e} there occurs a reflection at the repulsive core potential around $R = 0$ of an incoming density from infinity. The magnitude of $k$ determines the spread and velocity of the incoming and outgoing rings and the duration of the reflection dynamics. $c_5$ determines when the reflection occurs in time. In~\subref{fig:subfig4.1f} is shown the limit $k \to 0$ of potentials~\eqref{eqn:12.7} and~\eqref{eqn:12.1007}, the limiting behaviour of oscillations as the period goes to infinity. Note that the colouring is not to scale.}
\end{figure*}

Any nontrivial real solution of Eq.~\eqref{eqn:12.0} with the boundary
condition $\phi(\eta) \to 0$ as $\eta \to \infty$ is asymptotic to
$\nu \mathrm{Ai}(\eta)$ for some nonzero real constant $\nu$, where
$\mathrm{Ai}$ denotes the Airy function~\cite{NISTPII1}. The choice
$|\nu| = 1$ corresponds to the Hastings-McLeod
solution~\cite{springerlink:10.1007/BF00283254}, and the choice $|\nu|
< 1$ to the Segur-Ablowitz solution~\cite{Segur1981165}. Since by
Eqs.~\eqref{eqn:12.8c} and~\eqref{eqn:12.1008c} we have $\eta \geq 0$,
the solution is well approximated by taking the Airy function for
$\phi(\eta)$:
\begin{equation}
\label{eqn:12.1010}
\phi(\eta) = \nu \mathrm{Ai}(\eta).
\end{equation}
From Eqs.~\eqref{eqn:12.8},~\eqref{eqn:12.1008a},~\eqref{eqn:12.1008c}, and~\eqref{eqn:12.1010} it follows that $R |\psi|^2 \to 0$ for all $R$ as 
\begin{equation}
\label{eqn:12.1011}
T \to \frac{\left( l + \frac{1}{2}\right) \pi}{2 \sqrt{|k|}} - c_5,
\end{equation}
where $l \in \field{Z}$. We have made density plots corresponding to $\nu = 0.9$ and $c_4 = 3/2^\frac{3}{2}$ for various values of $k$ (see Fig.~\ref{Fig:4.1}). Note that unlike for the dark soliton-like solutions, here the solution is actually dynamical and evolving in time, describing a scattering (one-time or periodic) of the bright solution from the central potential.

The limiting potential with $k = 0$ can be obtained by choosing
\begin{align}
\label{eqn:k0.1}
c_1(T) &= c_4 (T + c_5)^{\frac{2}{3}}, \\
\label{eqn:k0.2}
\dot{c}_2(T) &= \frac{3}{2^{\frac{5}{2}} c_1(T)^3} - c_6.
\end{align}
Now Eq.~\eqref{eqn:12.8} with
\begin{align}
\label{eqn:k0.3}
\rho(R,T) &= \frac{R^{-\frac{1}{3}}}{c_4 (T + c_5)^{\frac{2}{3}}}, \\
\label{eqn:k0.4}
\varphi(R,T) &= \frac{1}{T+c_5}\left[\frac{R^2}{4}- c_6(T+c_5)^2 -\frac{3\sqrt{2}}{8c_4^3}\right]\\ \notag
&  + c_7,\\
\label{eqn:k0.5}
\eta(R,T) &= \frac{3}{2\sqrt{2}c_4} \left(\frac{R}{T + c_5}\right)^{\frac{2}{3}}
\end{align}
solves Eq.~\eqref{eqn:2} ($\sigma = 1$) with the potential
\begin{equation}
\label{eqn:k0.6}
V(R) = \frac{1}{9 R^2} + c_6.
\end{equation}
See Fig.~\ref{Fig:4.1}~\subref{fig:subfig4.1f}.

We note that the $k=0$ case is also obtained by choosing $c_1(T)= \mathrm{const.}$
and $\dot{c}_2(T) = \frac{3}{2^{\frac{5}{2}} c_1^3} - c_6$, which
leads to
\begin{align}
\rho(R) &= \frac{1}{c_1R^{\frac{1}{3}}}, \\
\varphi(T) &= \left(\frac{3}{2^{\frac{5}{2}} c_1^3} - c_6\right)T + c_7,\\
\eta(R) &= \frac{3}{2\sqrt{2}c_1} R^{\frac{2}{3}},\\
V(R) &= \frac{1}{9 R^2} + c_6.
\end{align}
In this case the equation may be called a ``nonlinear Bessel'' equation.

\section{Summary}\label{sec:summary}

We have obtained exact localised solutions of the radial
Gross-Pitaevskii equation describing cylindrically symmetric
systems. We have concentrated on solutions that may have
implications for studies of ring solitons in quantum gases, but
it should be noted that still other solutions are possible with
different choices of $h(\eta)$ and the integration ``constants''
$c_{1,2}(T)$.

The form of the potentials corresponding to exact solutions suggests that confinement of cold atoms in traps which are repulsive at the core should be studied further, especially in the context of ring soliton stability. Previous work and our own numerical simulations have shown that in general the ring dark soliton is ultimately destroyed by the snaking instability, and the ring collapses into a vortex necklace~\cite{PhysRevLett.90.120403}. Having a solution with a potential that has both a repulsive core, and a minimum at finite $R$, however, suggests that e.g. toroidal traps (also known as ring traps) might provide better stability for the ring dark soliton, and results of numerical investigations testing this idea will be reported in a later paper. Such potentials are experimentally feasible~\cite{Griffin2008,Baker2009,Sherlock2011}. We will also discuss elsewhere the possibility of creating dark solitons in toroidal traps using time-averaged potentials and/or adiabatic passage techniques~\cite{Garraway1998,Rodriguez2000,Martikainen2001,Harkonen2006}.

An additional aspect to such studies comes from the observation that some of our solutions allow one to choose freely the location of the dark or bright solution ($R_0$), as this location does not depend on the
parameters of the potential. Thus the long-time survival of the solution is not connected to locating the soliton-like structure at the minimum of some confining potential. It means that one may study ring soliton dynamics before the decay into vortices takes place, and look for similarities to such behavior in one-dimensional systems~\cite{PhysRevLett.84.2298}. Another feature is to consider the case of a
bright ring soliton in such traps, as it has not been studied very much in the past. Such considerations form the starting point for further work on ring solitons and soliton-like structures, including their stability and dynamics.

\begin{acknowledgments}
The authors thank the Finnish Cultural Foundation, Turun Suomalainen Yliopistoseura and Academy of Finland (project 133682) for funding this work.
\end{acknowledgments}

\bibliographystyle{apsrev4-1}
%

\end{document}